\documentclass[aps,prb,twocolumn,superscriptaddress]{revtex4-2}
\usepackage{graphicx}
\usepackage{bm}
\usepackage{xcolor}
\usepackage{amsmath}
\usepackage{amssymb}
\usepackage{hyperref}
\usepackage{tabularray}

\begin{document}

\title{Magnetic ground states in the kagome system YCu$_3$(OH)$_6$[(Cl$_x$Br$_{1-x}$)$_{3-y}$(OH)$_{y}$]}

\author{Aini Xu}
\affiliation{Beijing National Laboratory for Condensed Matter Physics, Institute of Physics, Chinese Academy of Sciences, Beijing 100190, China}
\affiliation{School of Physical Sciences, University of Chinese Academy of Sciences, Beijing 100190, China}
\author{Qinxin Shen}
\affiliation{Beijing National Laboratory for Condensed Matter Physics, Institute of Physics, Chinese Academy of Sciences, Beijing 100190, China}
\affiliation{School of Physical Sciences, University of Chinese Academy of Sciences, Beijing 100190, China}
\author{Bo Liu}
\affiliation{Beijing National Laboratory for Condensed Matter Physics, Institute of Physics, Chinese Academy of Sciences, Beijing 100190, China}
\affiliation{School of Physical Sciences, University of Chinese Academy of Sciences, Beijing 100190, China}
\author{Zhenyuan Zeng}
\affiliation{Beijing National Laboratory for Condensed Matter Physics, Institute of Physics, Chinese Academy of Sciences, Beijing 100190, China}
\affiliation{School of Physical Sciences, University of Chinese Academy of Sciences, Beijing 100190, China}
\author{Lankun Han}
\affiliation{Beijing National Laboratory for Condensed Matter Physics, Institute of Physics, Chinese Academy of Sciences, Beijing 100190, China}
\affiliation{School of Physical Sciences, University of Chinese Academy of Sciences, Beijing 100190, China}
\author{Liqin Yan}
\affiliation{Beijing National Laboratory for Condensed Matter Physics, Institute of Physics, Chinese Academy of Sciences, Beijing 100190, China}
\affiliation{School of Physical Sciences, University of Chinese Academy of Sciences, Beijing 100190, China}
\author{Jun Luo}
\affiliation{Beijing National Laboratory for Condensed Matter Physics, Institute of Physics, Chinese Academy of Sciences, Beijing 100190, China}
\author{Jie Yang}
\affiliation{Beijing National Laboratory for Condensed Matter Physics, Institute of Physics, Chinese Academy of Sciences, Beijing 100190, China}
\author{Rui Zhou}
\email{rzhou@iphy.ac.cn}
\affiliation{Beijing National Laboratory for Condensed Matter Physics, Institute of Physics, Chinese Academy of Sciences, Beijing 100190, China}
\author{Shiliang Li}
\email{slli@iphy.ac.cn}
\affiliation{Beijing National Laboratory for Condensed Matter Physics, Institute of Physics, Chinese Academy of Sciences, Beijing 100190, China}
\affiliation{School of Physical Sciences, University of Chinese Academy of Sciences, Beijing 100190, China}
\begin{abstract}
Quantum spin liquids represent exotic states of spin systems characterized by long-range entanglement and emergent fractionalized quasiparticles. 
It is generally believed that disorder is hostile to quantum spin liquids. In our study, we investigated the magnetic properties of a kagome system, YCu$_3$(OH)$_6$[(Cl$_x$Br$_{1-x}$)$_{3-y}$(OH)$_{y}$]. Within this system, some of the hexagons exhibit alternate bonds along the Cu-O-Cu exchange paths, while others remain uniform. We found that a long-range antiferromagnetic order emerges when uniform hexagons dominate. Conversely, a possible quantum-spin-liquid state arises when the number of alternate-bond hexagons exceeds about 2/3. Therefore, the alternate-bond hexagons, typically considered as disorders, actually serve as the building blocks of the quantum spin liquid in this system. Notably, the low-temperature properties of the quantum spin liquid are directly associated with the height of the out-of-plane yttrium ions, which may be linked to changes in superexchange energies. Our results suggest that understanding the magnetic ground states in this system lies beyond the theoretical framework of the Heisenberg model constructed on the kagome lattice.

\end{abstract}

\maketitle

\section{introduction}

Finding quantum spin liquids (QSLs) in real materials has been a long-standing challenge in condensed matter physics \cite{SavaryL17,ZhouY17,BroholmC20}. While there are many candidate materials that exhibit promising properties consistent with those of QSLs, these properties could have alternate explanations, many of which are associated with disorders. For example, strong disorders can destroy the long-range antiferromagnetic (AFM) order, resulting in non-QSL disordered states, such as random singlet states and valance-bond glass, which share some similar properties with QSLs \cite{BhattRN82,FisherDS94,LinYC03,TarziaM08,SongP21,KunduS20,MustonenOH22,LeeC23}. Some systems initially thought to host QSLs were later found to likely fall into the these categories \cite{PSinghRRP10,KawamuraH14,MustonenO18,UematsuK18,KimchiI18,LiuL18,LiuL20,WatanabeM18,HongW21,MaZ20,MaZ21}. Weak disorders can also lead to glassy properties or low-energy excitations that hinder the determination of ground states \cite{VriesMA08,FreedmanDE10,HanTH12,FuM15,HanTH16,KhuntiaP20,YYHuang2021,FengZL17,WeiYuan2017,FengZL18b,YingFu2021,WeiY21}. In rare cases, it has been proposed that disorders may enhance quantum entanglement and lead to QSLs \cite{SavaryL17b,SibilleR17,FurukawaT15,WenJJ17,SzirmaiP20}, but it is always an experimental question whether the ground states of real materials might simply be another manifestation of trivial paramagnetic states arising from randomness \cite{WatanabeK14,RiedlK19}. 

The kagome structure is one of the most important systems for searching for QSLs, since it possesses the most antiferromagnetically frustrated structure in two dimensions \cite{NormanMR16, YanS11,DepenbrockS12,JiangHC12,FanY12,RanY07,HeYC17,LiaoHJ17,HeYC14,GongSS15, WangYC18,GYSun2018}. Experimentally, many Cu$^{2+}$-based ($S$ = 1/2) kagome materials exhibit large exchange energies ($\sim$ 100 K) yet show no magnetic ordering, making them excellent candidates for QSLs. However, the nature of the ground states is often debated due to the presence of magnetic impurities, which typically dominate the low-energy spin excitations \cite{VriesMA08,FreedmanDE10,YYHuang2021,HanTH12,FuM15,HanTH16,KhuntiaP20,FengZL17,WeiYuan2017,FengZL18b,YingFu2021,WeiY21}. It has been found that the issue of weakly correlated impurity spins can be resolved by replacing the interlayer bivalent cations with a combination of Y$^{3+}$ and Cl$^{-}$/Br$^{-}$ ions \cite{SunW16,ChenXH20}. For YCu$_3$(OH)$_6$Cl$_3$ (YCu$_3$-Cl) with perfect Cu$^{2+}$ kagome planes, an AFM order is observed at $T_N \approx$ 15 K \cite{ZorkoA19,ZorkoA19b,BarthelemyQ19}. In the case where 1/9 of Cl$^{-}$ is replaced by (OH)$^{-}$, resulting in Y$_3$Cu$_9$(OH)$_{19}$Cl$_8$ (Y$_3$Cu$_9$-Cl), the lattice symmetry changes from $P\bar{3}m1$ to $R\bar{3}$; both AFM and disordered states have been reported \cite{PuphalP17,BarthelemyQ19,SunW21,HeringM22,ChatterjeeD23,WangJ23}. 

It has recently been discovered that there is no magnetic ordering in YCu$_3$(OH)$_6$Br$_2$[Br$_x$(OH)$_{1-x}$] (YCu$_3$-Br) \cite{ChenXH20,ZengZ22,LiuJ22,HongX22,LuF22,LiS24}, where Br$^{-}$ at the site on top of the yttrium ions is partially substituted by (OH)$^{-}$. Specific-heat measurements exhibit a quadratic temperature dependence at zero field and a linear term at high fields, consistent with the expected behavior of a Dirac QSL \cite{ZengZ22}. This is further supported by inelastic neutron scattering measurements, which reveal a conical spin continuum that could come from the convolution of two Dirac spinons \cite{ZengZ24}. Moreover, a one-ninth magnetization plateau is found in this material \cite{JeonS24}. Surprisingly, magnetic quantum oscillations are also observed around and above the one-ninth plateau \cite{ZhengG23}, demonstrating the presence of nontrivial excitations in this system. What sets YCu$_3$-Br apart from YCu$_3$-Cl is the large number of randomly distributed hexagons consisting of alternate bonds, which may significantly alter the low-energy properties \cite{LiuJ22,LiS24}. 

In YCu$_3$-Br, Y$^{3+}$ ions occupy the centers of the hexagons of the Cu$^{2+}$ kagome layers [Figs. \ref{fig1}(a) and \ref{fig1}(b)]. These ions can either reside within or outside the kagome planes due to the polar nature of (OH)$^-$ \cite{LiuJ22}. When Y$^{3+}$ is within the plane, the local structure remains undistorted, and all six Cu-O-Cu superexchange paths exhibit the same angle, as illustrated in Fig. \ref{fig1}(c). Conversely, if Y$^{3+}$ is out of plane, the Cu-O-Cu angle changes alternately [Fig. \ref{fig1}(d)]. These two types of hexagons are referred to as uniform hexagons (UHs) and alternate-bond hexagons (ABHs), denoted as $s$ and $n$, respectively  \cite{LiuJ22}, whose configurations of superexchanges are different as shown in Figs. \ref{fig1}(c) and \ref{fig1}(d). Given that YCu$_3$-Cl has a negligible percentage of ABHs while YCu$_3$-Br has a significantly higher count, it raises questions about whether the disordered state results from the randomness in the Cu-O-Cu superexchange paths associated with the ABHs. Furthermore, one might wonder if the proposed QSL state is once again overshadowed by disorders, leading to an ongoing debate akin to its sibling systems. 

In this work, we studied the magnetic properties of YCu$_3$(OH)$_6$[(Cl$_x$Br$_{1-x}$)$_{3-y}$(OH)$_y$] (YCu$_3$-ClBr). The proportion of ABHs, denoted as $n/(n+s)$, varies from nearly zero to approximately 0.7. We found that the AFM order and possible QSL emerge for $n/(n+s) \lesssim$ 1/3 and $\gtrsim$ 2/3, respectively. Mixed phases are observed in between these regions. Notably, the properties of the possible QSLs are directly associated with the height of the out-of-plane yttrium. Our results demonstrate that both the ABHs and the yttrium height play a critical role in the formation of the possible QSL in this system.

\section{experiments}

Single crystals of YCu$_3$-ClBr were grown using the hydrothermal method, similar to those reported previously \cite{ChenXH20,ZengZ22}. A mixture of Cu(NO$_3$)$_2$$\cdot$3H$_2$O (2.5 mmol) , Y(NO$_3$)$_3$$\cdot$6H$_2$O (5 mmol), KBr [15(1-x) mmol], KCl (15x mmol), and appropriate deionized water (from 1.2 to 1.8 mL) was put into  a hydrothermal synthesis autoclave reactor with the 25-mL polytetrafluoroethylene inner chamber. The autoclave was heated at 225 or 230 $^{\circ}$C for about a week. Attempts have also been made to add Ca or Mg into the pure Br samples but only less than 3\% of them are substituted at the yttrium site \cite{supp}. The studies here also include deuterated samples. The samples are denoted by the nominal chlorine content $x$, while the pure Br samples with D, Ca, and Mg will be denoted as 0\_D, 0\_Ca, and 0\_Mg, respectively. Additional labels of "a" and "b" are used if there are more than one batch of samples with different properties. All the crystals were further ultrasonically cleaned in water to remove possible impurities attached to the surfaces.  

Single-crystal x-ray diffraction (SCXRD) was used to determine the crystal structure. The electron spin resonance (ESR) experiments were carried out with a JEOL JES-FA200 ESR spectrometer at X-band frequencies with microwave power of 1 mW at 9.06 GHz . Specific-heat and magnetic-susceptibility measurements were performed on a physical property measurement system (Quantum Design)  and a magnetic property measurement system (Quantum Design), respectively. Nuclear magnetic resonance (NMR) measurements were conducted using a phase-coherent pulsed NMR spectrometer. To obtain sufficient NMR signal intensity, the $^{79,81}$Br-NMR spectra were measured on a collection of several single crystals by adding Fourier transforms of the spin-echo signal recorded at regularly spaced frequency values. We stacked the single-crystal flakes along the $c$ direction, ensuring that the applied magnetic field was aligned with the $c$ axis. The $T_1$ relaxation time was measured using the saturation recovery method and determined by fitting to the theoretical curve. We emphasize that the properties of samples can vary from batch to batch, even when using the same growth conditions. Therefore, it is essential to take different measurements from the same batch of samples to ensure consistent results. 

\section{results}

\begin{figure}[tbp]
\includegraphics[width=\columnwidth]{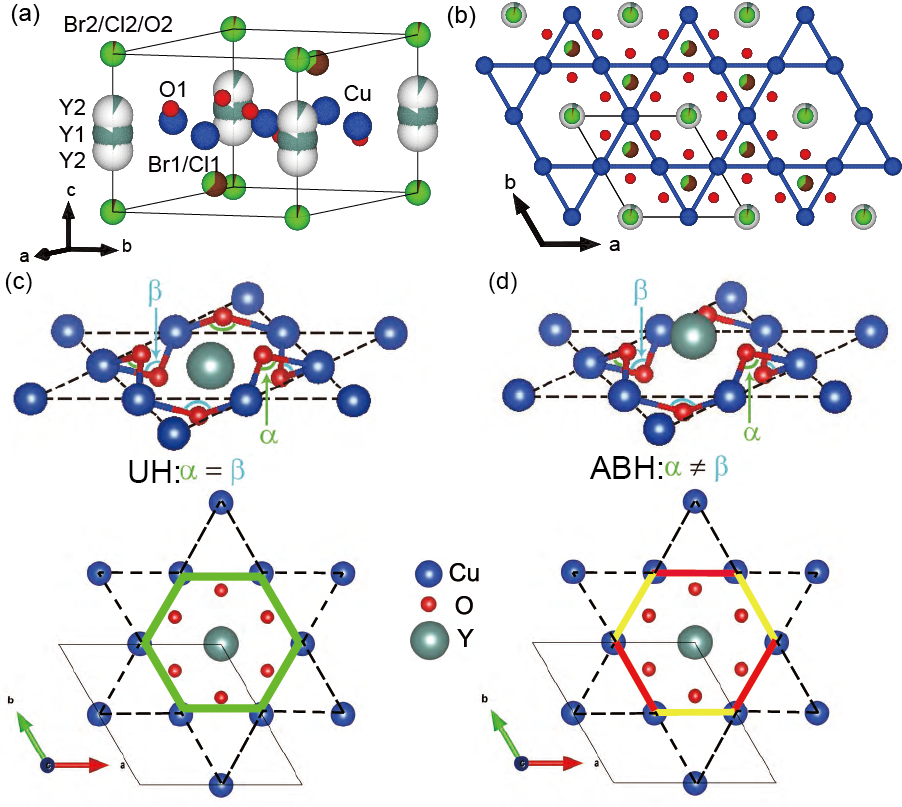}
 \caption{(a) and (b) Schematic diagram of the crystal structure of YCu$_3$-ClBr ($x$ = 0.21) in side and top views, respectively, where hydrogens are not shown. The crystal structures of other samples are the same but with different occupancies at Y$^{3+}$ and Br$^-$ sites. The black solid lines represent the unit cell.  (c) The schematic diagram of the local crystal structure of YCu$_3$-ClBr with a Y$^{3+}$ ion positioned at the center of the Cu hexagon. Only Cu, O, and Y atoms are shown. In the upper panel, the Cu-O-Cu angles associated with the hexagon are labeled as $\alpha$ and $\beta$ for the oxygen atoms above and below the kagome plane, respectively. The superexchange bonds within the hexagon are depicted as green lines in the lower panel. (d) Similar to (a) but with the Y$^{3+}$ situated outside the kagome plane. The alternate bonds within the hexagon are represented by the red and yellow lines in the lower panel. }
 \label{fig1}
\end{figure}

We first give an overall picture of the crystal structure of YCu$_3$-ClBr determined from SCXRD, as illustrated in Figs. \ref{fig1}(a) and \ref{fig1}(b). The detailed information of the structure is given in the Supplemental Material \cite{supp}. For all the samples, the lattice structure maintains the $P\bar{3}m1$ space group, which means that the kagome planes formed by Cu$^{2+}$ ions remain undistorted. The Cl$^-$ and Br$^-$ ions can occupy both the Br1 and Br2 positions. Part of the Br2 position is also occupied by the oxygen associated with the (OH)$^-$ ions. Therefore, the value of $y$ in the molecular formula YCu$_3$(OH)$_6$[(Cl$_x$Br$_{1-x}$)$_{3-y}$(OH)$_y$] is only associated with the replacement of Cl$^-$/Br$^-$ at the Br2 position, but we keep this formula for simplicity and consistency with our previous results \cite{ShivaramBS24}. 

As pointed out previously \cite{LiuJ22}, due to the polarization of the (OH)$^-$ ion, the Y$^{3+}$ ion sitting at the center of the hexagons will be pulled out of the kagome planes. This results in two different local crystal structures as shown in Figs. \ref{fig1}(c) and \ref{fig1}(d). In the first case labeled as UH, the Y$^{3+}$ ion sits within the kagome plane and the value of $\alpha$ for the three Cu-O-Cu angles above the plane is the same as that of $\beta$ for those below the plane. The nearest superexchanges are thus supposed to be the same. In the second case labeled as ABH, the Y$^{3+}$ ion sits out of the kagome plane and therefore $\alpha \neq \beta$, resulting in alternate superexchange bonds within the hexagon.

\begin{figure}[tbp]
\includegraphics[width=\columnwidth]{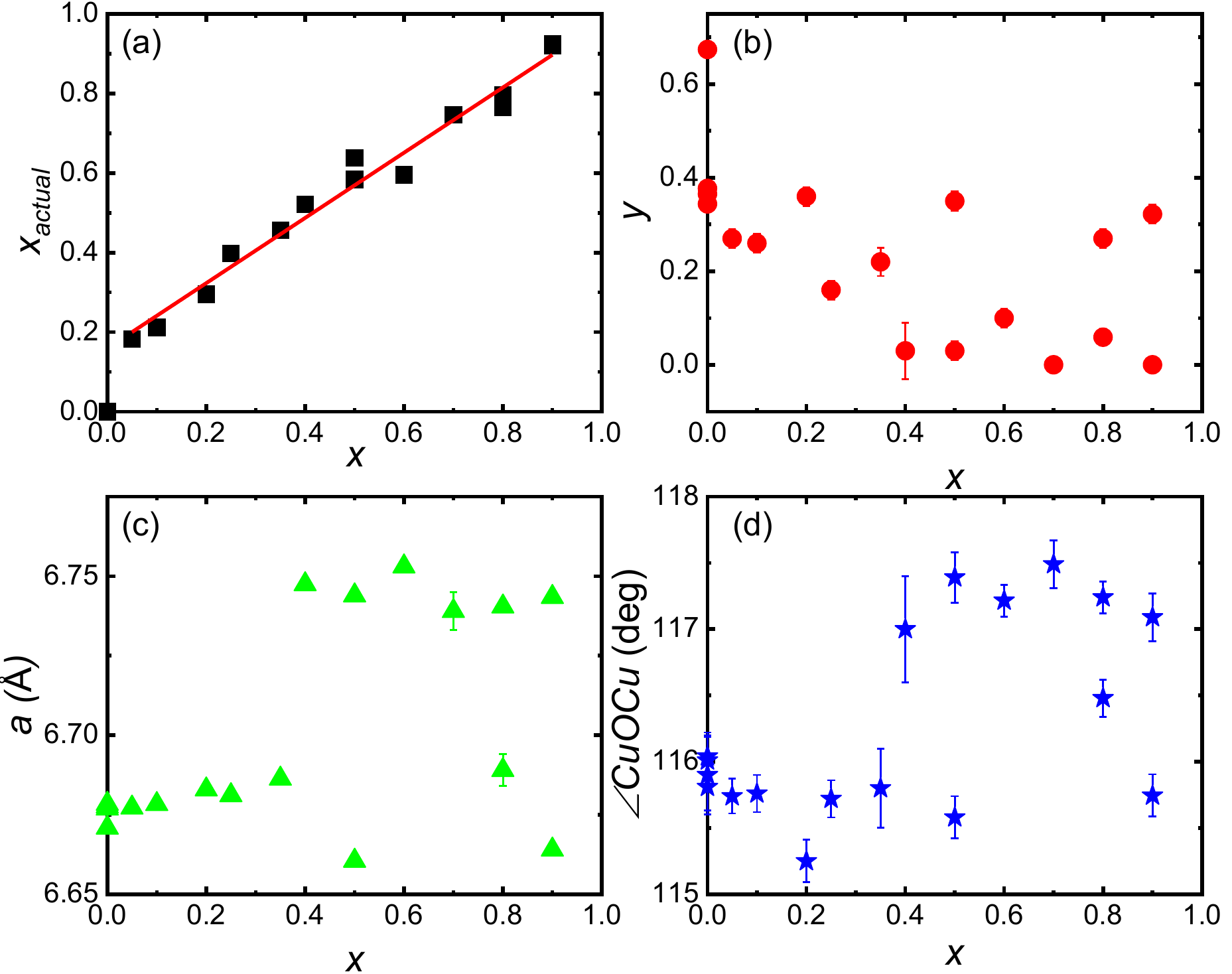}
 \caption{(a) Relationship between the nominal chlorine content $x$ and the actual content $x_{actual}$. The solid line is the linear fitting result without the $x$ = 0 point. (b)-(d) The $x$ dependence of $y$, lattice parameter $a$, and Cu-O-Cu angles.}
 \label{fig2}
\end{figure}

Figure \ref{fig2}(a) shows the change of the actual chlorine content $x_{actual}$ as the function of the nominal chlorine content $x$. Except for the $x$ = 0 sample, where $x_{actual}$ apparently  is zero, $x_{actual}$ changes linearly with $x$ as $x_{actual}$ = 0.16 + 0.82$x$. For the value of $y$, i.e., the content of (OH)$^-$ ions at the Br2 position, there is no clear $x$ dependence. Even for the same $x$, the values of $y$ may be very different, as shown in Fig. \ref{fig2}(b). Figures \ref{fig2}(c) and \ref{fig2}(d) show the $x$ dependence of the lattice constant $a$ and the Cu-O-Cu angle, respectively. Again, these values can be very different even for the same $x$. This means that the value of $x$ is NOT a proper variable to describe the change of the lattice parameters and the physical properties as shown later. Note that the value of $y$ is directly associated with the occupancy of the Y$^{3+}$ ion at the Y2 position \cite{LiuJ22}, which can be determined more precisely in the structural refinement, suggesting that the occupancy of the Y$^{3+}$ ion at the Y2 position may provide a consistent description of the structural changes.

\begin{figure}[tbp]
\includegraphics[width=\columnwidth]{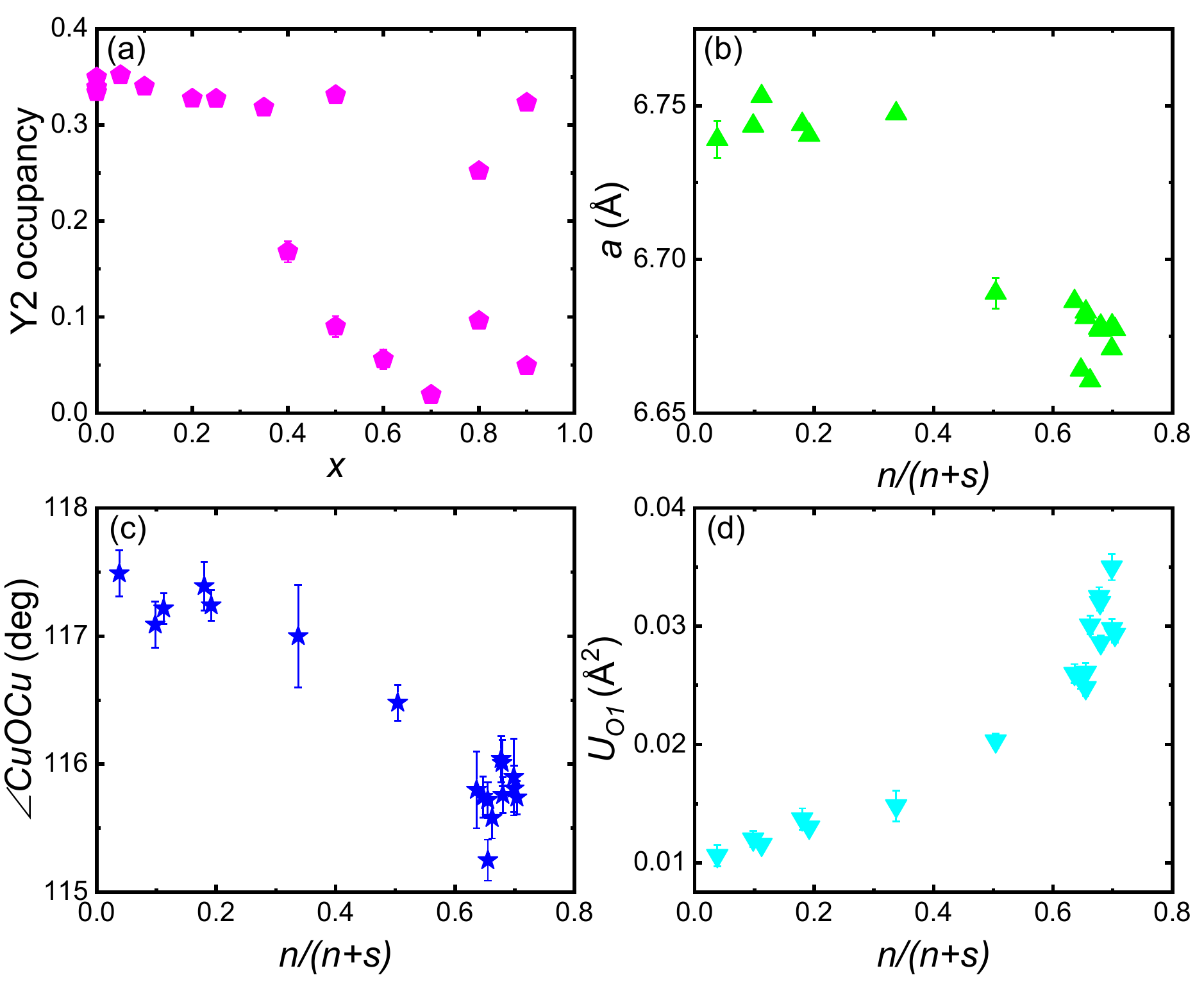}
 \caption{(a) $x$ dependence of Y2 occupancy. (b)-(d) The $n/(n+s)$ dependence of the lattice constant $a$,  the Cu-O-Cu angle, and the atomic displacement parameter U of oxygen at the O1 position. }
 \label{fig3}
\end{figure}

Figure \ref{fig3}(a) shows the $x$ dependence of Y2 occupancy. Noting that this value is equal to half of $n/(n+s)$, where $n$ and $s$ are the number of ABHs and UHs, respectively. The change of lattice constant $a$ with $n/(n+s)$ is plotted in Fig. \ref{fig3}(b). The value of $a$ changes smoothly with $n/(n+s)$ and becomes smaller with less number of ABHs, which is more or less apparent since the deviation of Y$^{3+}$ ions from the kagome plane will shrink the in-plane lattice. In the mean time, the increasing number of Y2 occupancy also results in an expansion along the $c$ axis \cite{supp}. Figure \ref{fig3}(c) shows the $n/(n+s)$ dependence of the Cu-O-Cu angle, which also shows a monotonous change. It should be pointed out that we only consider one O1 position associated with the Cu-O-Cu angle in the refinement [Fig. \ref{fig1}(a)], but there are actually two of them for the ABHs as shown in Fig. \ref{fig1}(d). While it is impossible to refine the SCXRD data with a certain mixture of one and two oxygen positions, the isotropic atomic displacement parameter  $U_{O1}$ at the O1 position monotonously increases with the increase of $n/(n+s)$ [Fig. \ref{fig3}(d)], resulting from the increasing number of ABHs with two O1 positions. Therefore, the change of the Cu-O-Cu angle with $n/(n+s)$ should not be simply understood as the change of the angle but rather an increase of ABHs with two distinct Cu-O-Cu angles. It is interesting to note that when $n/(n+s)$ is close to 2/3, $U_{O1}$ increases rapidly, which is associated with the increase of Y$^{3+}$ height at the Y2 position.

\begin{figure}[tbp]
\includegraphics[width=\columnwidth]{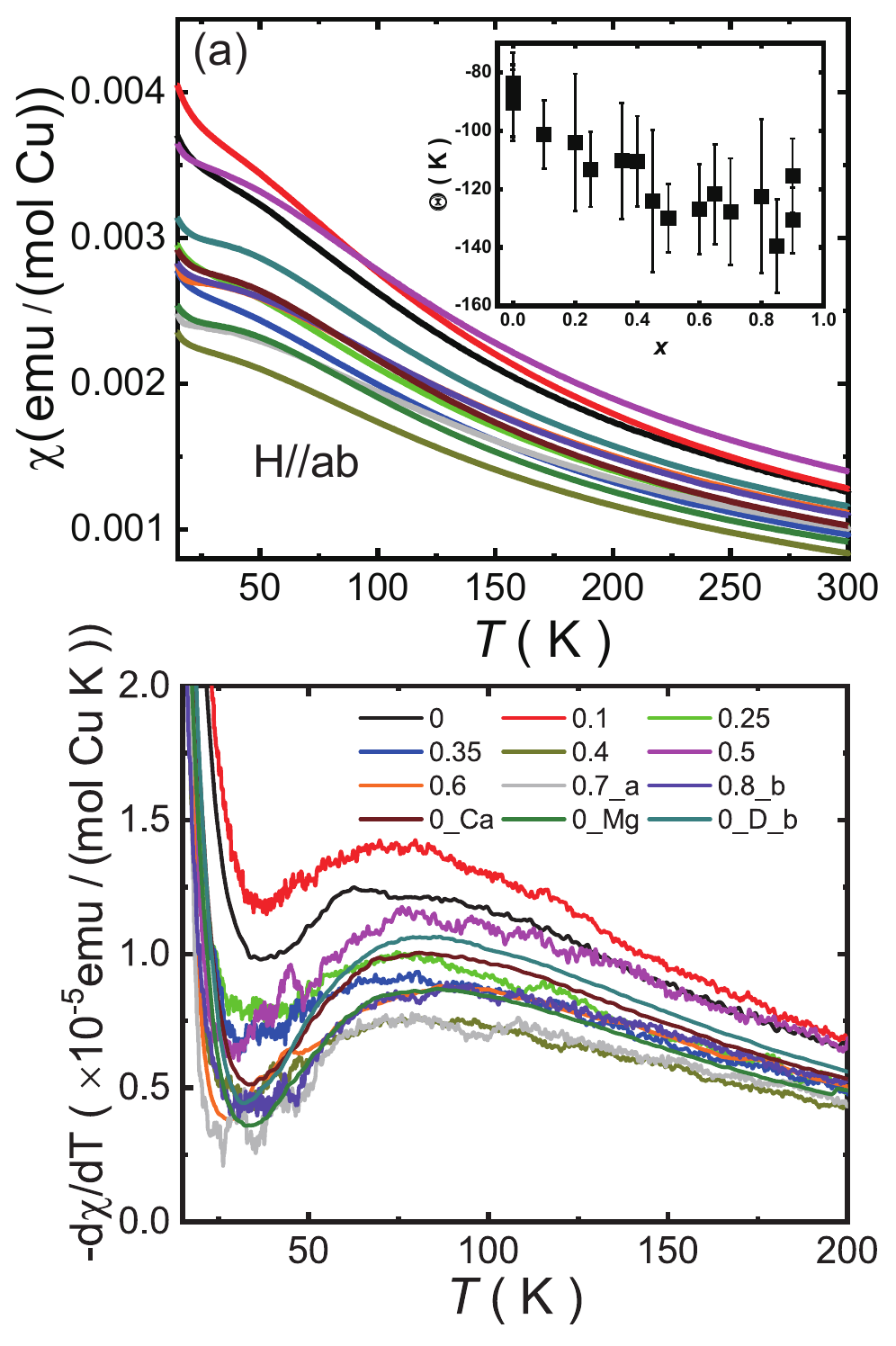}
 \caption{(a) Temperature dependence of the magnetic susceptibility $\chi$ measured at 1000 Oe for $T >$ 20 K. The samples cover a range of $x$ values from 0 to 0.9, although they are not labeled for simplicity. The inset shows the $x$ dependence of the Weiss temperature $\Theta$, which is averaged from ZFC and FC processes using different fitting temperature ranges. The error bars represent standard deviations obtained by fitting different temperature ranges. (b) The temperature dependence of $-d\chi/dT$.  }
 \label{fig4}
\end{figure}

Figure \ref{fig4}(a) shows the temperature dependence of the magnetic susceptibility $\chi$ for $T>$ 20 K. All the samples exhibit similar behaviors. The high-temperature data can be well fitted using the Curie-Weiss function $\chi$ = $C/(T-\Theta)+\chi_0$, where $C$, $\Theta$ and $\chi_0$ are the Curie constant, Weiss temperature, and a temperature-independent background, respectively. The effective moments calculated from $C$ are all close to the value of the $S$=1/2 magnetic moment with $g \approx$ 2. As shown in the inset of Fig. \ref{fig4}(a), $|\Theta|$ increases with increasing $x$ up to $x \approx$ 0.5, after which it tends to saturate. In the intermediate temperature range  (approximately 20 to 50 K), the rate of increase of $\chi$ with decreasing temperature decelerates. This behavior is more pronounced in the temperature dependence of -$d\chi/dT$ where a dip emerges below 40 K [Fig. \ref{fig4}(b)]. We note that no substantial differences are observed among different samples. However, as shown below, the low-temperature properties of these samples reveal distinct ground states, including AFM order, possible QSL, and mixed phases.

\begin{figure}[tbp]
\includegraphics[width=\columnwidth]{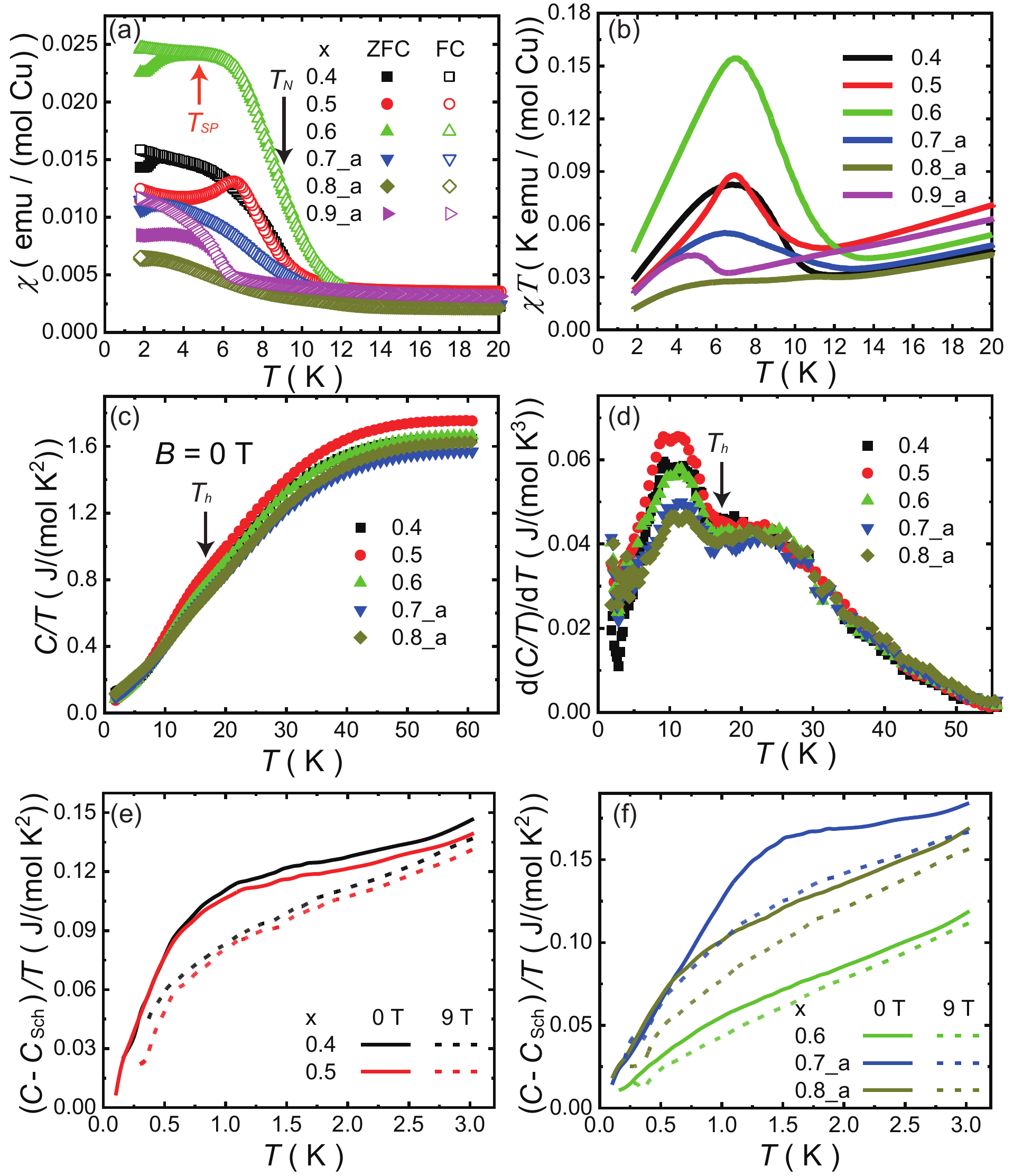}
 \caption{Magnetic susceptibility and specific heat results of AFM ordered samples. (a) The temperature dependence of $\chi$ at low temperatures measured at 1000 Oe. The red and black arrows indicate the splitting temperature $T_{SP}$ between the ZFC and FC processes and $T_N$, respectively, for the $x$ = 0.6 sample. (b) The temperature dependence of $\chi T$ for the FC process. (c), (d) The temperature dependence of $C/T$ and $d(C/T)/dT$ at zero field, respectively. The arrows indicate $T_h$ as discussed in the main text. (e), (f) The temperature dependence of the low-temperature $C/T$ with the nuclear Schottky anomaly subtracted for the $x$ = 0.4/0.5 and 0.6/0.7\_a/0.8\_a samples, respectively.}
 \label{fig5}
\end{figure}

We first present the results of AFM ordered samples in Fig. \ref{fig5}. For these samples, the magnetic susceptibility $\chi$ increases rapidly with decreasing temperature below approximately 11 K, as shown in Fig. \ref{fig5}(a). Figure \ref{fig5}(b) illustrates the temperature dependence of $\chi T$, revealing that this upturn is faster than $1/T$. Below a temperature denoted as $T_{SP}$, the zero-field-cooling (ZFC) and field-cooling (FC) magnetic susceptibilities become different. These features closely resemble those observed in YCu$_3$-Cl \cite{SunW16,ZorkoA19}. It is important to note that various methods exist for determining the AFM transition temperature $T_N$ from magnetic-susceptibility data, and the resulting values may differ. In our study, we define $T_N$ as the temperature at which the increase of $\chi$ is most rapid, as shown in Fig. \ref{fig5}(a), which is consistent with the NMR results presented later.

\begin{figure}[tbp]
\includegraphics[width=\columnwidth]{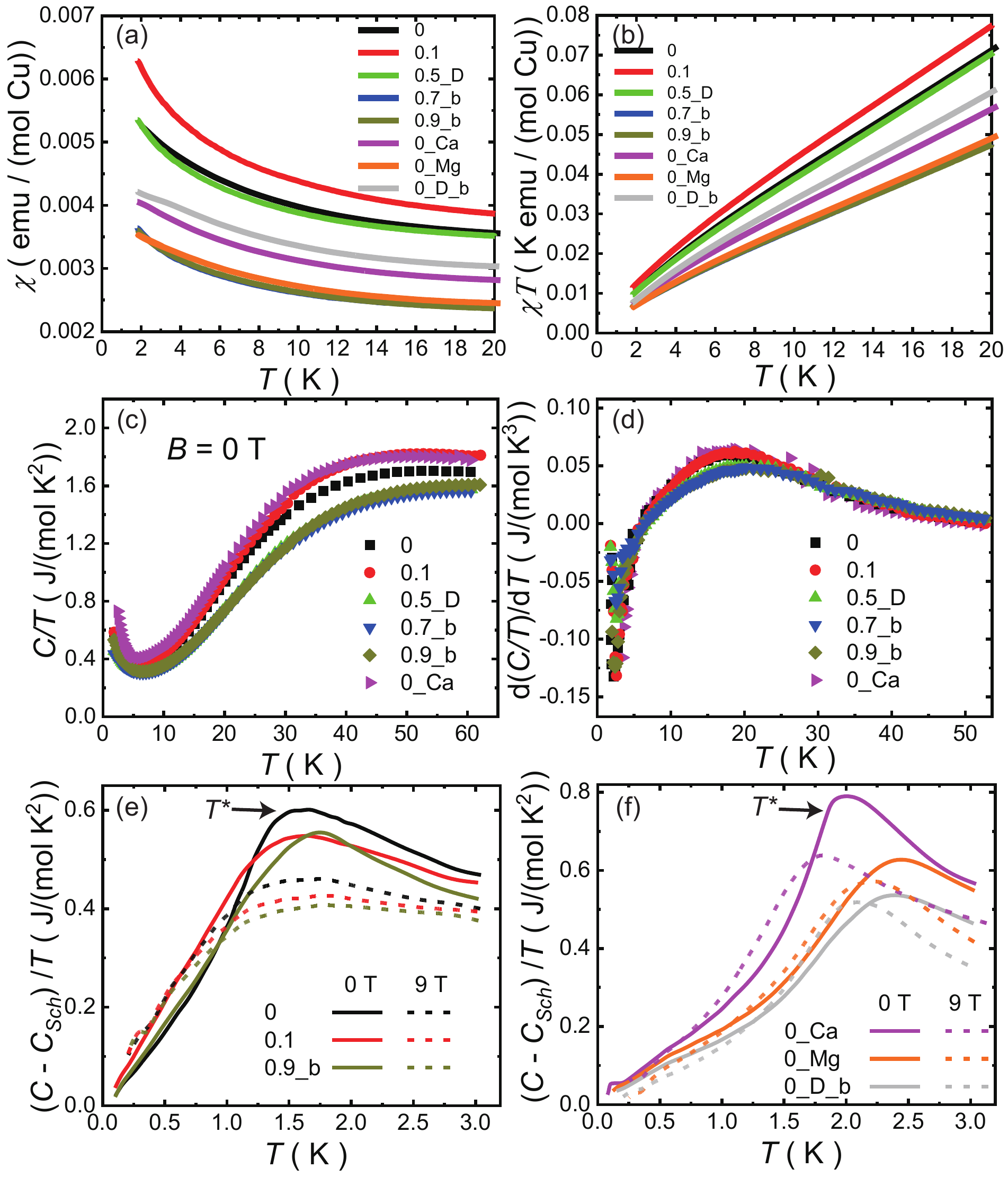}
 \caption{Magnetic susceptibility and specific heat results of possible QSL samples. (a) The temperature dependence of $\chi$ at low temperatures measured at 1000 Oe. (b) The temperature dependence of $\chi T$ at low temperatures. (c), (d) The temperature dependence of $C/T$ and $d(C/T)/dT$ at zero field, respectively.  (e), (f) The temperature dependence of the low-temperature $C/T$ with the nuclear Schottky anomaly subtracted for two sets of samples, respectively. The arrows indicate $T^*$ as discussed in the main text.}
 \label{fig6}
\end{figure}

Figure \ref{fig5}(c) shows the temperature dependence of the specific heat $C/T$ at zero field for the AFM ordered samples. No anomaly is observed at $T_N$, indicating the AFM transition is weak. Instead, a broad hump appears at about 16 K, similar to the behavior observed in YCu$_3$-Cl with AFM order \cite{ZorkoA19}. This feature becomes more pronounced in the derivative $d(C/T)/dT$, which exhibits rapid increase as the temperature approaches the hump temperature $T_h$, as shown in Fig. \ref{fig5}(d). Figures \ref{fig5}(e) and \ref{fig5}(f) present the low-temperature $C/T$, where the nuclear Schottky anomaly (in the form of A$T^{-3}$) has been subtracted, as previously reported \cite{ZengZ22}. Although the zero-field specific heat follows a $T^2$ temperature dependence at very low temperatures, i.e., $C/T$ = $\alpha T$, applying a magnetic field of 9 T suppresses the specific heat. This suggests that the $T^2$ temperature dependence of the specific heat likely arises from trivial two-dimensional spin fluctuations.

\begin{figure}[tbp]
\includegraphics[width=\columnwidth]{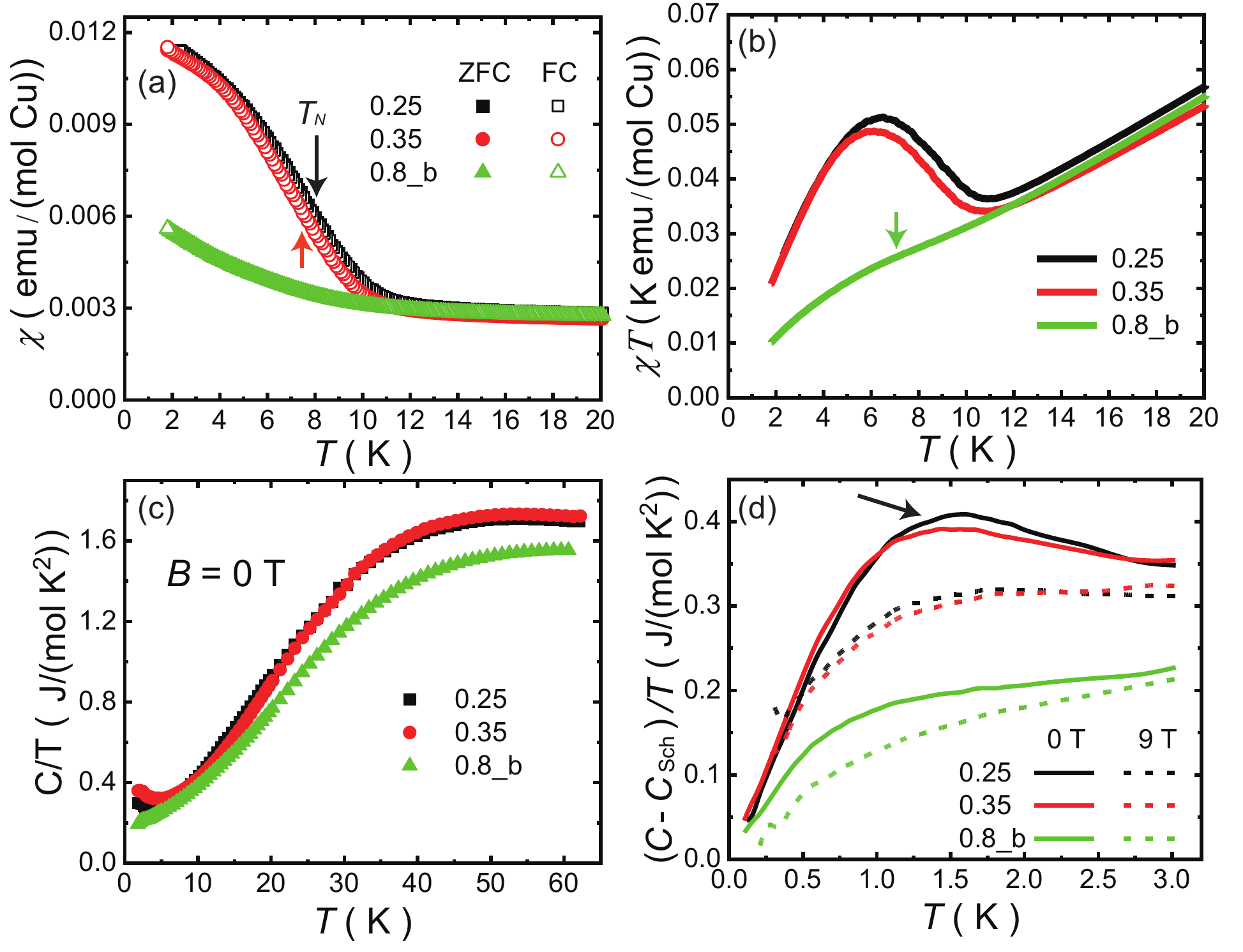}
 \caption{Magnetic susceptibility and specific heat results of the samples with mixed phases. (a) The temperature dependence of $\chi$ at low temperatures measured at 1000 Oe. (b) The temperature dependence of $\chi T$ for the FC process. The arrows indicate $T_N$ for the $x$ = 0.25 and 0.35 samples. (c) The temperature dependence of $C/T$ at zero field. The arrow indicate $T_N$ for the $x$ = 0.8\_b sample. (d) The temperature dependence of the low-temperature $C/T$ with the nuclear Schottky anomaly subtracted.The arrow indicates $T^*$ as discussed in the main text.}
 \label{fig7}
\end{figure}

Next, we present the results for the samples in the possible QSL states in Fig. \ref{fig6} including those of YCu$_3$-Br \cite{ZengZ22}. The magnetic susceptibilities of all samples show a slight upturn at low temperatures [Fig. \ref{fig6}(a)], but no FC and ZFC difference is observed. The disappearance of the AFM order becomes more evident in the temperature dependence of $\chi T$ [Fig. \ref{fig6}(b)], where its value continuously decreases with decreasing temperature. Meanwhile, the hump in the specific heat at about 16 K also vanishes, as shown in Figs. \ref{fig6}(c) and \ref{fig6}(d). Intriguingly, a new hump emerges at much lower temperature denoted as $T^*$ [Figs. \ref{fig6}(e) and \ref{fig6}(f)], suggesting an accumulation of entropy at low temperatures. The specific heat characteristics of the 0.1 and 0.9\_b samples resemble those of the $x$ = 0 sample. Specially,  they exhibit a linear temperature dependence of $C/T$ at zero field and a positive zero-K intercept under the magnetic field, as shown in Fig. \ref{fig6}(e), i.e., $C$ = $\gamma(H) T + \alpha T^2$ with $\gamma(H) >$ 0. It is worth noting that the specific heat values for these samples at low temperatures are approximately four times larger than those observed in the ordered samples [Figs. \ref{fig5}(e) and \ref{fig5}(f)]. However, for the possible QSL samples in Fig. \ref{fig6}(f), the low-temperature  ( $<$ 1 K ) specific heat values become significantly smaller. Most notably, no positive zero-K intercept of $C/T$ is found under a magnetic field of 9 T. These results suggest that although these samples are also in the possible QSL states, they exhibit different properties from those of the samples in Fig. \ref{fig6}(e).

\begin{figure}[tbp]
\includegraphics[width=\columnwidth]{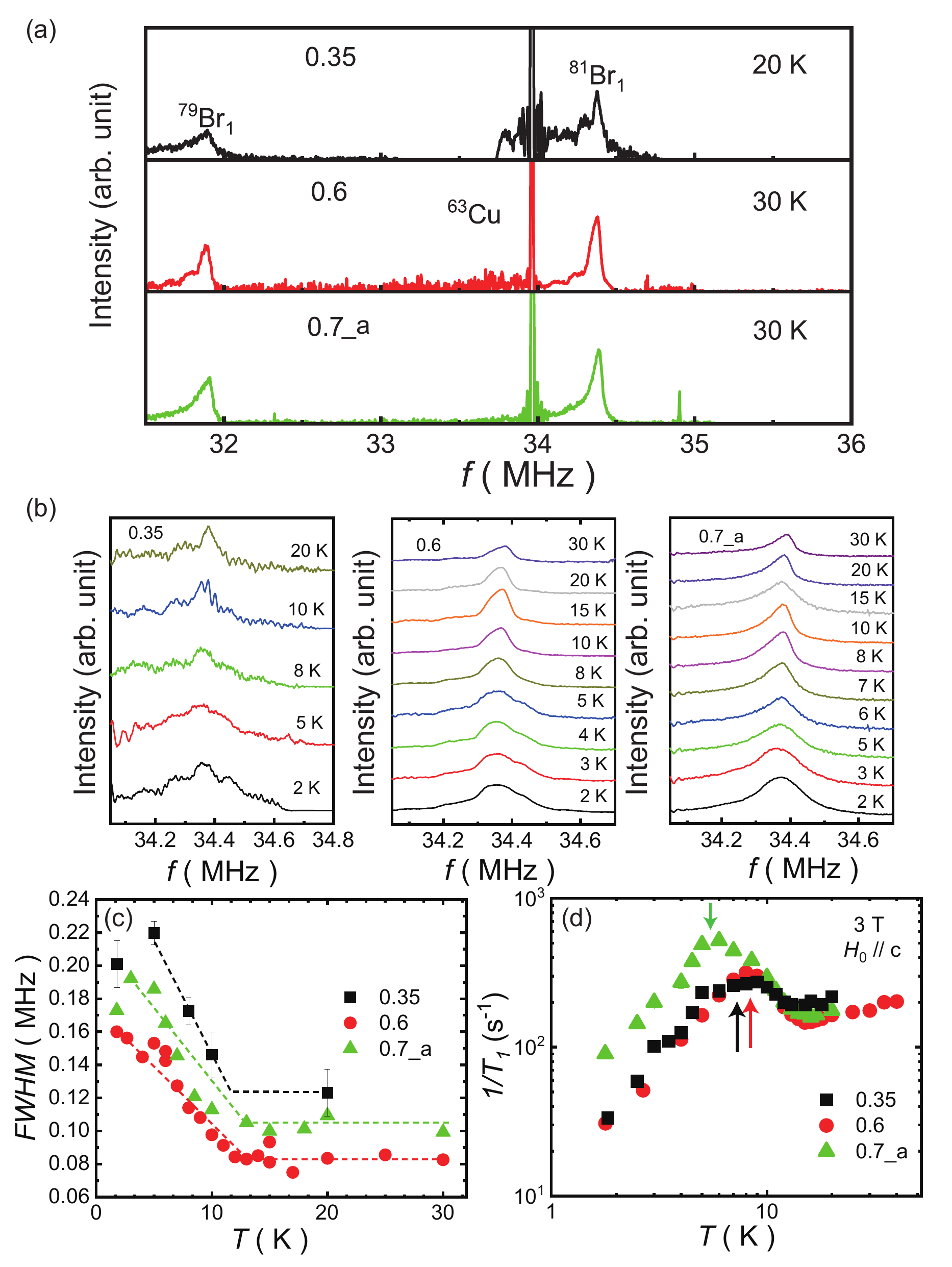}
 \caption{NMR results at 3 T with the field along the $c$ axis. (a) NMR spectra at 20 or 30 K. (b) NMR spectra for the $^{81}Br$ at various temperatures for the 0.35 (left), 0.6 (middle), and 0.7\_a (right) samples. (c) Temperature dependence of the FWHM of $^{81}$Br NMR spectra. The dashed lines are guides to the eye. (d) Temperature dependence of the spin-lattice relaxation rate 1/$T_1$ of $^{81}$Br. The arrows indicate $T_N$.}
 \label{fig8}
\end{figure}

There are samples that exhibit behaviors distinct from both the AFM and possible QSL samples. Figures \ref{fig7}(a) and \ref{fig7}(b) show the temperature dependence of $\chi$ and $\chi T$ for these samples, respectively. While the upturn that signifies the AFM transition is present, there is no discernible difference between ZFC and FC processes. For the $x$ = 0.8\_b sample, while the temperature dependence of $\chi$ looks similar to those possible QSL samples, the temperature dependence of $\chi T$ clearly reveals the appearance of the magnetic transition [Fig. \ref{fig7}(b)]. The high-temperature hump in $C/T$ also disappears, as shown in Figs. \ref{fig7}(c). At low temperatures, the specific heat of the $x$ = 0.25 and 0.35 samples behaves similar to the possible QSL samples in Fig. \ref{fig6}(e), but the values are smaller, as illustrated in Fig. \ref{fig7}(d). The hump at $T^*$ still exists although it becomes less significant. For the $x$ = 0.8\_b sample, the specific heat resembles that of the AFM ordered samples [Figs. \ref{fig5}(e) and \ref{fig5}(f)] and is significantly suppressed under the 9 T field.

\begin{figure}[tbp]
	\includegraphics[width=\columnwidth]{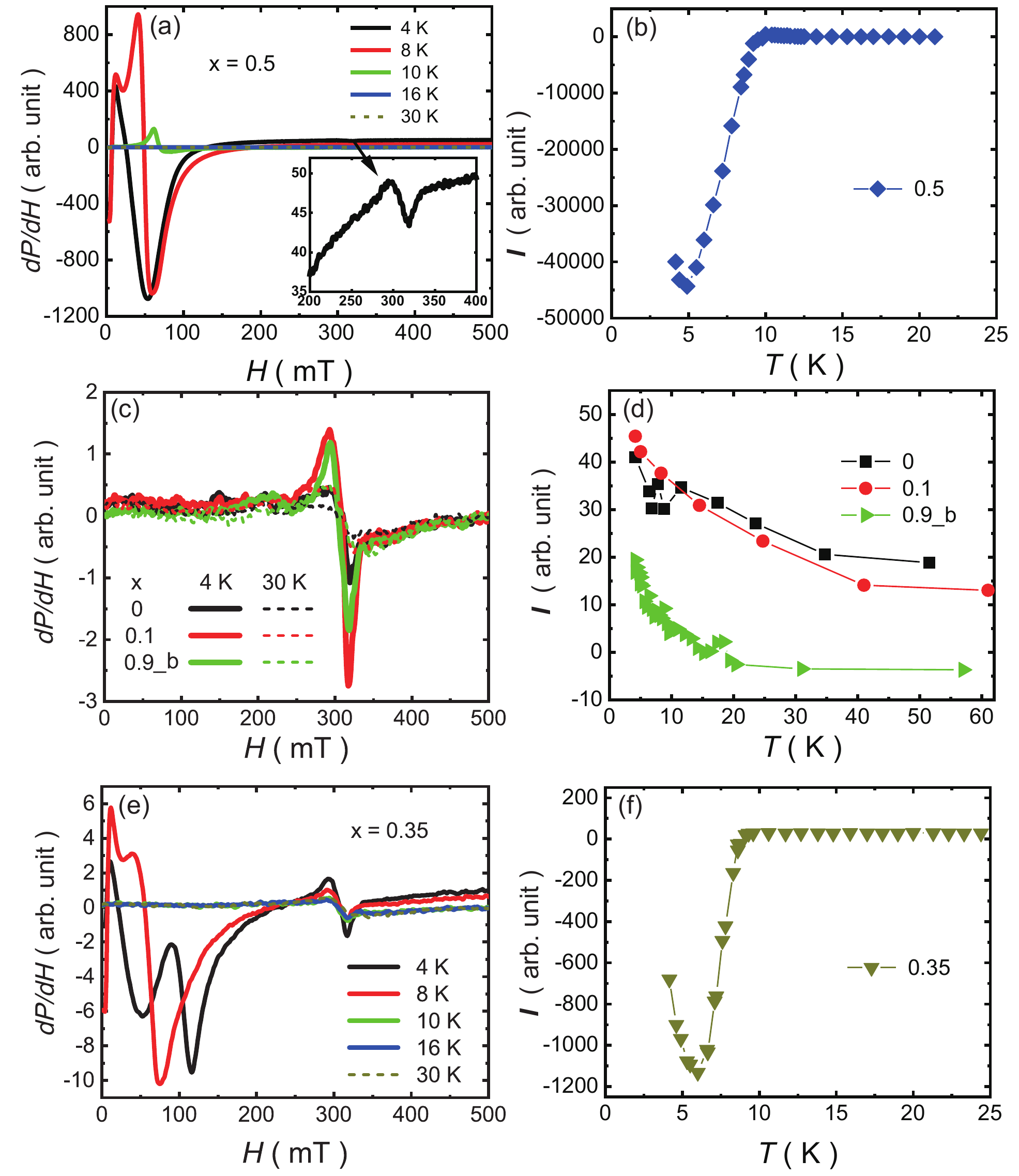}
 \caption{ESR results. (a) ESR signals of the AFM ordered $x$ = 0.5 sample at various temperatures. (b) Temperature dependence of the integrated intensity from 0 to 200 mT for the $x$ = 0.5 sample. (c) ESR signals of $x$ = 0, 0.1, and 0.9\_a sample at 4 and 30 K. (d) Temperature dependence of the integrated intensity from 0 to 200 mT for the $x$ = 0, 0.1, and 0.9\_a sample. (e) ESR signals of the $x$ = 0.35 sample at various temperatures. (f) Temperature dependence of the integrated intensity from 0 to 200 mT for the $x$ = 0.35 sample.}
 \label{fig9}
\end{figure}

The existence of the AFM order is further investigated using the NMR technique. Figure \ref{fig8}(a) shows the NMR spectra at 3 T and temperatures of 20 or 30 K for the $x$ = 0.35, 0.6, and 0.7\_a samples, all of which are in the paramagnetic state. There are two Br$^-$ sites, namely Br1 and Br2, existing above the Cu$^{2+}$ triangles and hexagons, respectively \cite{ZengZ22}. Since Cl$^-$ preferentially replaces Br2 atoms \cite{supp}, only the signal from the Br1 site is observed for $x$ $\geq$ 0.35 samples, or $x_a \gtrsim$ 0.5 where $x_a$ is the actual chlorine content. As temperature decreases, the knight shift $^{81}K$ exhibits negligible changes for all samples. Rather than line splitting, only line broadening is observed even in the AFM ordered samples with $x$ = 0.6 and 0.7\_a [Fig. \ref{fig8}(b)]. Figure \ref{fig8}(c) illustrates the temperature dependence of the full width at half maximum (FWHM) for three samples. While the high-temperature values of the $x$ = 0.35 sample exceed those of the $x$ = 0.6 and 0.7\_a samples, the broadening of the peaks at low temperatures are rather the same. Evidence for the establishment of the AFM orders in the $x$ = 0.6 and 0.7\_a samples comes from the temperature dependence of the spin-lattice relaxation rate 1/$T_1$, as shown in Fig. \ref{fig8}(d). For the $x$ = 0.6 and 0.7\_a samples, the value of 1/$T_1$ decreases with decreasing temperature below the transition temperature $T_N$, which is due to the suppression of spin fluctuations and a typical feature of the AFM order \cite{ZhouR13}. However, it is important to note that no splitting is observed in the spectra, indicating that the order is rather glassy in nature. Surprisingly, the behavior of the 1/$T_1$ in the $x$ = 0.35 sample exhibits similar behaviors with those in the $x$ = 0.6 and 0.7\_a samples although the peak feature becomes less prominent . This suggests the presence of the AFM order in this sample although its low-temperature specific heat behaves like the possible QSL samples [Fig. \ref{fig7}(d)]. 

A natural explanation for the results in the $x$ = 0.35 sample is that both the AFM and possible QSL phases exist in this sample. This phase coexistence can be further elucidated through the ESR measurements. Figure \ref{fig9}(a) shows the ESR signals of the AFM ordered $x$ = 0.5 sample. At 30 K, only the paramagnetic signal at about 320 mT (with $g \approx$ 2) is detectable. As temperature decreases, various peaks emerge below about 200 mT, indicating the effects of the AFM order. The temperature dependence of the integrated intensity from 0 to 200 mT exhibits a transition at about 9 K [Fig. \ref{fig9}(b)], consistent with the $T_N$ value determined from other methods. The ESR signals of the possible QSL samples are dominated by the paramagnetic signal across all temperatures [Fig. \ref{fig9}(c)]. Additionally, the temperature dependence of the integrated intensity below 200 mT still displays an enhancement at low temperatures, which may come from the ferromagnetic (FM) clusters in these samples \cite{ShivaramBS24}. The behavior of the $x$ = 0.35 sample closely resembles that of the $x$ = 0.5 sample, but the amplitude of the low-field signals is approximately two orders of magnitude smaller, as shown in Figs. \ref{fig9}(e) and \ref{fig9}(f), implying the existence of AFM clusters in the $x$ = 0.35 sample. We note that the differences of the NMR results between the $x$ = 0.35 and 0.6 samples, including the temperature dependence of the FWHM and 1/$T_1$ [Figs. \ref{fig8}(c) and \ref{fig8}(d)], are not very significant . This discrepancy arises because the hyperfine coupling in NMR mainly depends on the relative position between the Br1 site and the neighboring Cu$^{2+}$ moments, whereas the low-field ESR signal originates from the FM components of the AFM clusters. In the $x$ = 0.35 sample, these FM components become very weak, as evidenced by the ZFC and FC magnetic susceptibilities. 

\begin{figure}[tbp]
\includegraphics[width=\columnwidth]{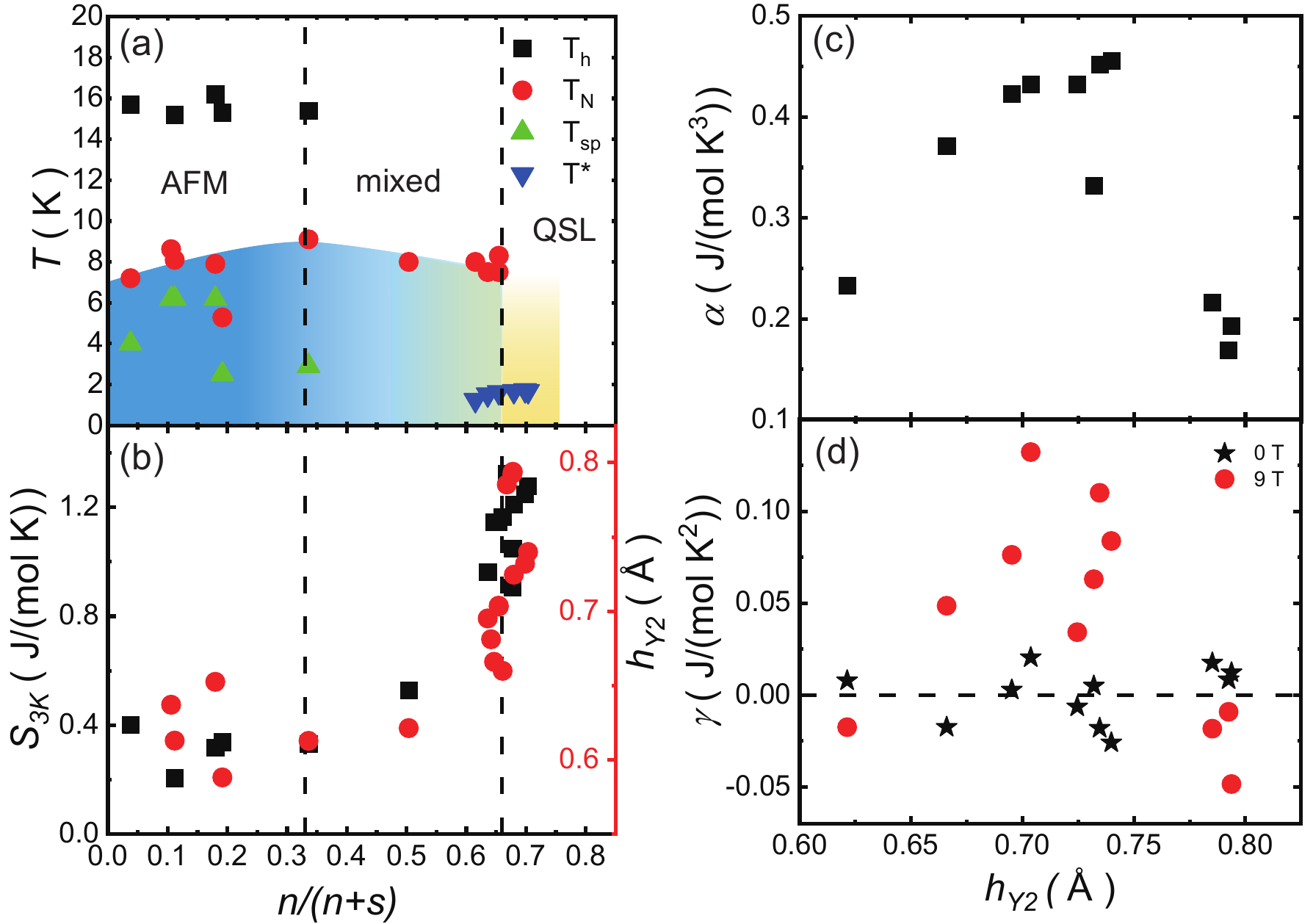}
 \caption{(a) Magnetic phase diagram of the YCu$_3$-ClBr system with the value of $n/(n+s)$ as the variable. (b) The $n/(n+s)$ dependence of the entropy below 3 K and the height of the out-of-plane yttrium, denoted as $S_{3K}$ and $h_{Y2}$, respectively. (c), (d) The $h_{Y2}$ dependence of the $T^2$ coefficient $\alpha$ and the $T$-linear coefficient $\gamma$ at 0 and 9 T for the possible QSL and mixed-phases samples with $n/(n+s)$ around 2/3.}
 \label{fig10}
\end{figure}

Although we have denoted the samples with the chlorine content, it is clear that the magnetic ground states could not solely be determined by it. Different ground states can be observed in the same chlorine-substituted samples. This is in line with the $x$ dependence of the structural parameters (Fig. \ref{fig2}). Therefore, one would expect that the physical properties of this system should be also better described by $n/(n+s)$, i.e., the proportion of ABHs (Fig. \ref{fig3}). Figure \ref{fig10}(a) shows the magnetic phase diagram of the YCu$_3$-ClBr system. All the AFM ordered samples are located at $n/(n+s) \lesssim$ 1/3, while those with the possible QSL states have $n/(n+s) \gtrsim$ 2/3. Mixed phases of the AFM and possible QSL states are found in between. Since this phase diagram is based on the bulk properties, we cannot rule out the possibilities that there might exist small magnetic clusters in the possible QSL samples \cite{ShivaramBS24}, or small quantum-disordered clusters in the AFM samples. Notably, AFM-ordered samples are observed only when the actual content of chlorine exceeds approximately 0.5, and no sign of mixed phase is found for the chlorine content less than about 0.2. This is likely due to the different chemical properties between Cl$^-$ and Br$^-$ ions. The latter appears to facilitate the substitution of (OH)$^-$ ions at the positions connecting two Y$^{3+}$ ions in adjacent kagome planes, resulting in an increased abundance of ABHs in the sample.  

Figure \ref{fig10}(b) illustrates the change in entropy below 3 K as a function of $n/(n+s)$. The value of $n/(n+s)$ exhibits significant variation around $n/(n+s)$ = 2/3. This observation suggests that the formation of the possible QSL and its associated properties are not solely determined by the proportion of ABHs. Upon examining detailed information about the lattice structure, we find that the height of out-of-plane yttrium, denoted as $h_{Y2}$, behaves similarly to the low-temperature entropy, as depicted in Fig. \ref{fig10}(b). Figures \ref{fig10}(c) and \ref{fig10}(d) present the $h_{Y2}$ dependence of the $T^2$ coefficient $\alpha$ and $T$-linear coefficient at 0 and 9 T in the specific heats for the possible QSL samples and the mixed-phase samples exhibiting QSL behaviors. Both exhibit a dome-shaped behavior, with the largest values of $\alpha$ and $\gamma_{9T}$ occurring at approximately $h_{Y2} \approx$ 0.7 \AA. 

\section{discussions}

Previous theoretical studies have highlighted the significance of ABHs in the formation of the possible QSL in YCu$_3$-Br \cite{LiuJ22}. Our results unambiguously provide experimental evidence that leads to further understanding of the importance of ABHs. The three regions in Fig. \ref{fig10}(a) clearly demonstrate that the AFM order emerges when the amount of UHs dominates [$n/(n+s) \lesssim$ 1/3], while the possible QSL state arises when $n/(n+s) \gtrsim$ 2/3. Interestingly, the value of $2/3$ closely aligns with the site percolation threshold of the kagome lattice, i.e. 1-2sin($\pi$/18). This alignment may explain the disappearance of the AFM order for $n/(n+s) \gtrsim$ 2/3. Consequently, there likely exist different sizes of AFM clusters in the mixed-phases region, contributing to the seemingly AF transitions observed in the magnetic susceptibility and 1/$T_1$ from the NMR measurements [Figs. \ref{fig7}(b) and \ref{fig8}(d)]. Furthermore, even in the well ordered AFM samples, small clusters of disordered phase may persist due to the presence of ABHs, as suggested by the broadening (but not splitting) of NMR spectra (Fig. \ref{fig8}) and the persistent spin dynamics in YCu$_3$-Cl \cite{ZorkoA19}. Similarly, the possible QSL samples may also contain small magnetic clusters, likely due to the existence of clustered UHs, as indicated by the low-field ESR signal [Fig. \ref{fig9}(d)] and the nonlinear field dependence of magnetization \cite{ShivaramBS24}. It is worth noting that fluctuating magnetic droplets have also been found in another QSL candidate NaYbSe$_2$ \cite{ZhuZ23}, although their origin may differ. 

In AFM-ordered samples, the specific heat exhibits a hump above $T_N$ and there is a splitting between the ZFC and FC magnetic susceptibilities below $T_N$. The hump has been attributed to the presence of the Dzyaloshinskii-Moriya (DM) interaction \cite{ArhT20}, which can also lead to the canting of the magnetic moments, resulting in the ZFC and FC splitting. These features disappear in the mixed-phases samples, suggesting the weakening or randomness of the DM interaction in the presence of ABHs. This is in line with theoretical studies that suggest the DM interaction is hostile to the formation of the QSLs \cite{ArhT20}. 

In the possible QSL samples, the behavior of the low-temperature specific heat varies. In some samples, the specific heat follows the expectation of the Dirac QSLs, i.e., $C$ = $\gamma(H) T + \alpha T^2$ with $\gamma(H) >$ 0. However, in other samples, $\gamma(H)$ approaches zero or becomes negative. This distinction is highlighted in Figs. \ref{fig10}(c) and \ref{fig10}(d), where the underlying cause is linked to $h_{Y2}$.  Since $h_{Y2}$ should be related to the Cu-O-Cu angles [as depicted in Fig. \ref{fig1}(d)], it is reasonable to hypothesize that the low-energy excitations of the possible QSL samples are associated with the changes in superexchange interactions. Unfortunately, the SCXRD technique employed in this study lacks the necessary resolution to precisely identify the detailed information regarding the various Cu-O-Cu angles.  Further investigations are warranted to unravel this issue.

Our results clearly demonstrate that the proposed QSL in YCu$_3$-ClBr is not overshadowed by bond disorders. On the contrary, these disorders and randomness may play a crucial role in forming the QSL state. In a sense, the ABHs should not be treated as disorders, as they constitute approximately two-thirds of the bonds in the possible QSL samples. While there have been some theoretical studies on the role of alternate bonds, our results raise questions about some of them. For example, certain attempts have introduced significant differences in exchange energies between alternating bonds or strong disorders to explain the shoulder like feature observed between approximately 20 and 50 K in magnetic susceptibility or Knight shift \cite{LiuJ22,LiS24}. However, contrary to theoretical predictions, this feature does not exhibit sensitivity to the amount of ABHs, as illustrated in Fig. \ref{fig1}(d). Therefore, measuring magnetic susceptibility or Knight shift does not appear to be an effective probe for the QSL state at low fields. This lack of sensitivity could be attributed to the existence of FM clusters \cite{ShivaramBS24}. More appropriate theoretical models are thus needed to account for our results.

It is interesting to note that in the Y$_3$Cu$_9$-Cl system, where $n/(n+s)$ is exactly 2/3 but the kagome plane is distorted, different reports have revealed either AFM or disordered phases \cite{PuphalP17,BarthelemyQ19,SunW21,ChatterjeeD23}. The reported value of $h_{Y2}$ for this system falls between 0.62 and 0.67 \AA, which would result in a change from the AFM to possible QSL state in our samples [Fig. \ref{fig10}(a)]. Despite having different crystal structure, there appear to be important similarities between these two systems. A theoretical model consisting of 1/3 of UHs and 2/3 of ABHs has been shown to exhibit a (1/3, 1/3) AFM order \cite{HeringM22}, which was experimentally verified in Y$_3$Cu$_9$-Cl \cite{ChatterjeeD23,WangJ23}. This model further suggests that a classical spin liquid could exist with proper values of superexchanges. Notably, the wavevectors of the spin excitations in YCu$_3$-Br \cite{ZengZ24} are the same as those in the (1/3,1/3) AFM order \cite{ChatterjeeD23}. It is thus intriguing to explore whether a QSL may emerge from this theoretical model. Overall, our results suggest that achieving a QSL state in the YCu$_3$-ClBr system may require a theoretical framework beyond the Heisenberg kagome model.

\section{conclusion}

In conclusion, we demonstrate that the YCu$_3$-ClBr system serves as an excellent platform to study the possible Dirac QSL. The presence of approximately 2/3 ABHs is found to be crucial for establishing the proposed QSL. Furthermore, the establishment of the possible QSL is accompanied by a substantial increase in entropy at low temperatures. The low-energy excitations of the possible QSL are associated with the height of the out-of-plane yttrium, which may be connected to changes of the superexchange energies. These results emphasize the important role played by alternating bonds and promote the development of a new theoretical framework to understand the QSL physics in this system.

\acknowledgments

S. L. thanks Professor Z. Y. Meng, Professor Z. Liu, and Professor Y. Zhou for discussions. This work is supported by the National Key Research and Development Program of China (Grants No. 2022YFA1403400, No. 2021YFA1400400, and No. 2023YFA1406100) and the Strategic Priority Research Program(B) of the Chinese Academy of Sciences (Grants No. XDB33000000 and No. GJTD-2020-01). A portion of this work was carried out at the Synergetic Extreme Condition User Facility (SECUF).

\end{document}